\documentclass[12pt]{article}
\usepackage[utf8]{inputenc}
\usepackage{authblk}
\usepackage{setspace}
\usepackage[margin=1.25in]{geometry}
\usepackage{graphicx}
\usepackage{subcaption}
\usepackage{amsmath}
\usepackage{float}

\usepackage[backend=bibtex,style=nejm, 
citestyle=numeric-comp,
sorting=none]{biblatex}

\begin{document}
\newcommand{\cD}{\ensuremath{\mathcal D} }
\newcommand{\scrD}{\ensuremath{\mathscr D} }
\newcommand{\cM}{\ensuremath{\mathcal M} }
\newcommand{\cN}{\ensuremath{\mathcal N} }
\newcommand{\cO}{\ensuremath{\mathcal O} }
\newcommand{\Pbb}{\ensuremath{\mathbb P} }
\newcommand{\cP}{\ensuremath{\mathcal P} }
\newcommand{\That}{\ensuremath{\widehat{T}} }
\newcommand{\Tc}{\ensuremath{T_{\text{c}}} }
\newcommand{\Thatc}{\ensuremath{\That_{\text{c}}} }
\newcommand{\al}{\ensuremath{\alpha} }
\newcommand{\be}{\ensuremath{\beta} }
\newcommand{\ga}{\ensuremath{\gamma} }
\newcommand{\Ga}{\ensuremath{\Gamma} }
\newcommand{\de}{\ensuremath{\delta} }
\newcommand{\De}{\ensuremath{\Delta} }
\newcommand{\eps}{\ensuremath{\epsilon} }
\newcommand{\la}{\ensuremath{\lambda} }
\newcommand{\lalat}{\ensuremath{\la_{\text{lat}}} }
\newcommand{\muhat}{\ensuremath{\widehat{\mu}} }
\newcommand{\mulat}{\ensuremath{\mu_{\text{lat}}} }
\newcommand{\si}{\ensuremath{\sigma} }
\newcommand{\om}{\ensuremath{\omega} }
\newcommand{\nn}{\nonumber }
\newcommand{\gsim}{\ensuremath{\gtrsim} }
\newcommand{\lsim}{\ensuremath{\lesssim} }
\newcommand{\SO}[1]{\ensuremath{\text{SO(}#1\text{)}} }
\newcommand{\X}{\ensuremath{\!\times\!} }
\newcommand{\chidof}{\ensuremath{\mbox{$\chi^2/\text{d.o.f.}$}}}
\newcommand{\pf}{\ensuremath{\text{pf}\,} }
\newcommand{\Tr}[1]{\ensuremath{\text{Tr}\left[ #1 \right]} }
\newcommand{\vev}[1]{\ensuremath{\left\langle #1 \right\rangle} }
\newcommand{\eq}[1]{Eq.~(\ref{#1})}
\newcommand{\fig}[1]{Fig.~\ref{#1}}
\newcommand{\tab}[1]{Table~\ref{#1}}
\newcommand{\secref}[1]{Sec.~\ref{#1}}
\newcommand{\refcite}[1]{Ref.~\cite{#1}}
\newcommand{\crit}{\vert_{\text{crit.}}}

\title{Deconfinement Phase Transition in Bosonic BMN Model at General Coupling}
\author[1*$\dagger$]{Navdeep Singh Dhindsa}
\author[1,2]{Anosh Joseph}
\author[3]{Abhishek Samlodia}
\author[4]{David Schaich}
\affil[1]{Department of Physical Sciences, Indian Institute of Science Education and Research - Mohali, Knowledge City, Sector 81, SAS Nagar, Punjab 140306, India}
\affil[2]{National Institute for Theoretical and Computational Sciences, School of Physics and Mandelstam Institute for Theoretical Physics, University of the Witwatersrand, Johannesburg, Wits 2050, South Africa}
\affil[3]{Department of Physics, Syracuse University, Syracuse, New York 13244, United States}
\affil[4]{Department of Mathematical Sciences, University of Liverpool, Liverpool L69 7ZL, United Kingdom}
\affil[*]{navdeep.s.dhindsa@gmail.com}
\affil[$\dagger$]{Contribution to the proceedings of the XXV
  DAE-BRNS HEP Symposium 2022, 12-16 December 2022, IISER Mohali, India}

\onehalfspacing
\date{}
\maketitle

\begin{abstract}

We present our analysis of the deconfinement phase transition in the bosonic BMN matrix model. The model is investigated using a non-perturbative lattice framework. We used the Polyakov loop as the order parameter to monitor the phase transition, and the results were verified using the separatrix ratio. The calculations are performed using a large number of colors and a broad range of temperatures for all couplings. Our results indicate a first-order phase transition in this theory for all the coupling values that connect the perturbative and non-perturbative regimes of the theory.

\end{abstract}

\section{Introduction}

The main goal of this analysis is to explore the dependence of the critical transition temperature ($\hat T_c$) of the bosonic BMN model on a deformation parameter ($\hat \mu$). The lattice action of the model is obtained by discretizing it on a lattice with $N_\tau$ sites and is given as
\begin{eqnarray}
S_{\text{lat}} &=& - \frac{N}{4 \lalat} \sum_{n = 0}^{N_\tau - 1} \mbox{Tr} \Bigg[ \sum_{i = 1}^9 \left(\cD_+ X_i\right)^2 + \frac{1}{2} \sum_{i<j; i, j = 1}^9 \left[X_i, X_j\right]^2 \nonumber \\ 
&& + \sum_{I = 1}^3 \left( \frac{\mulat}{3} X_I \right)^2 + \sum_{A = 4}^9 \left( \frac{\mulat}{6} X_A \right)^2 - \sum_{I, J, K = 1}^3 \frac{\sqrt{2} \mulat}{3} \eps_{IJK} X_I X_J X_K \Bigg].
\end{eqnarray}
The model has nine scalars, $X_i$ and a gauge field is realized through the covariant derivative. The finite-difference operator acting on a scalar has the form: 
\begin{equation}
    \cD_+ X_i(n) \equiv U(n) X_i(n + 1) U^{\dag}(n) - X_i(n),
\end{equation}
with $U(n)$ denoting the gauge link field attached to the site $n$. The lattice action is formulated using dimensionless parameters ($\mulat \equiv a \mu$, and $\lalat \equiv a^3 \lambda$, with $a$ denoting the lattice spacing). An observable we use to find the phase transition in the model is the susceptibility, $\chi$, of the Polyakov loop magnitude $|P|$, which is 
\begin{equation}
    \chi \equiv N^2 \left(\vev{|P|^2} - \vev{|P|}^2 \right).
\end{equation}

Here are several parameters that are of particular relevance:
\begin{align}
  \That & \equiv \frac{T}{\la^{1/3}} = \frac{1}{N_{\tau}\lalat^{1/3}}, &
  \muhat & \equiv \frac{\mu}{\la^{1/3}} = \frac{\mulat}{\lalat^{1/3}}, &
  \frac{\That}{\muhat} & = \frac{T}{\mu} = \frac{1}{N_{\tau} \mulat}.
\end{align}
\begin{figure}[h!]
  \centering
    \includegraphics[width=0.32\linewidth]{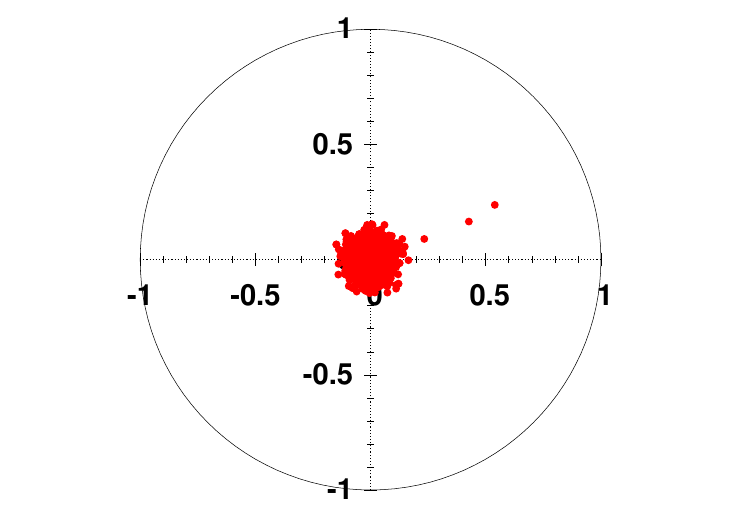}
    \includegraphics[width=0.32\linewidth]{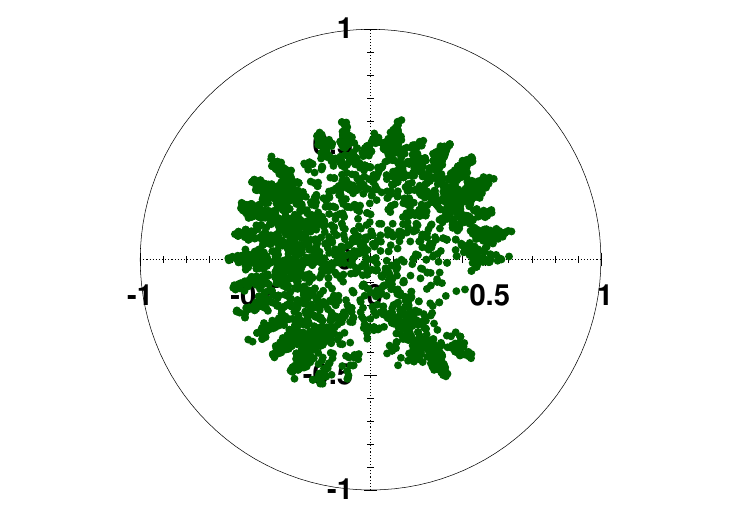}
    \includegraphics[width=0.32\linewidth]{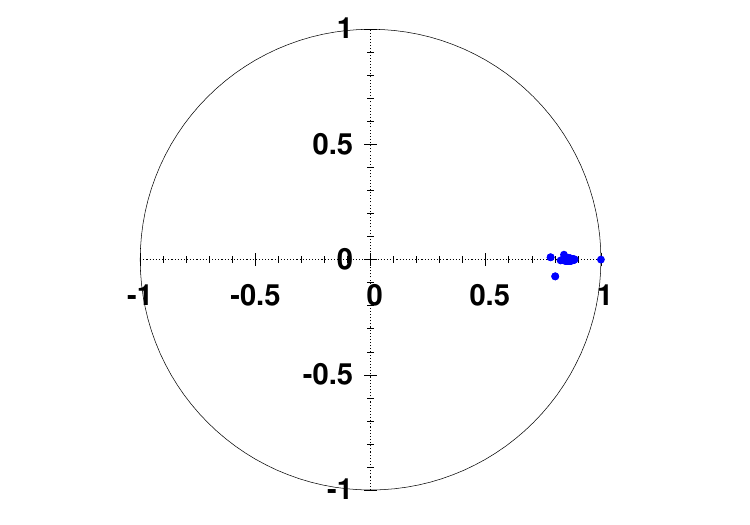}
    \includegraphics[width=0.32\linewidth]{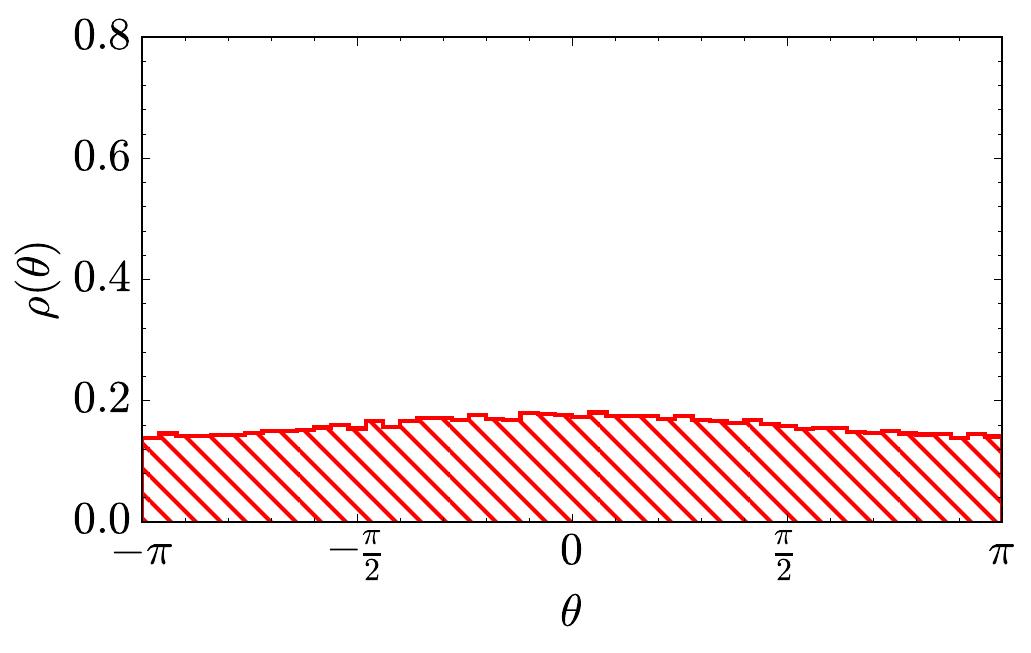}
    \includegraphics[width=0.32\linewidth]{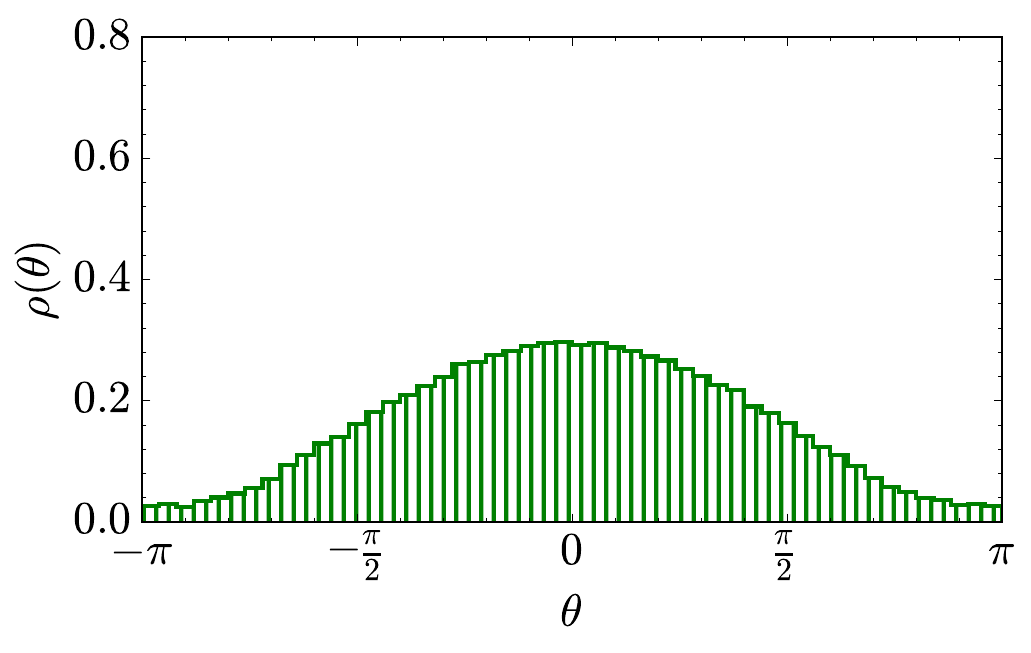}
    \includegraphics[width=0.32\linewidth]{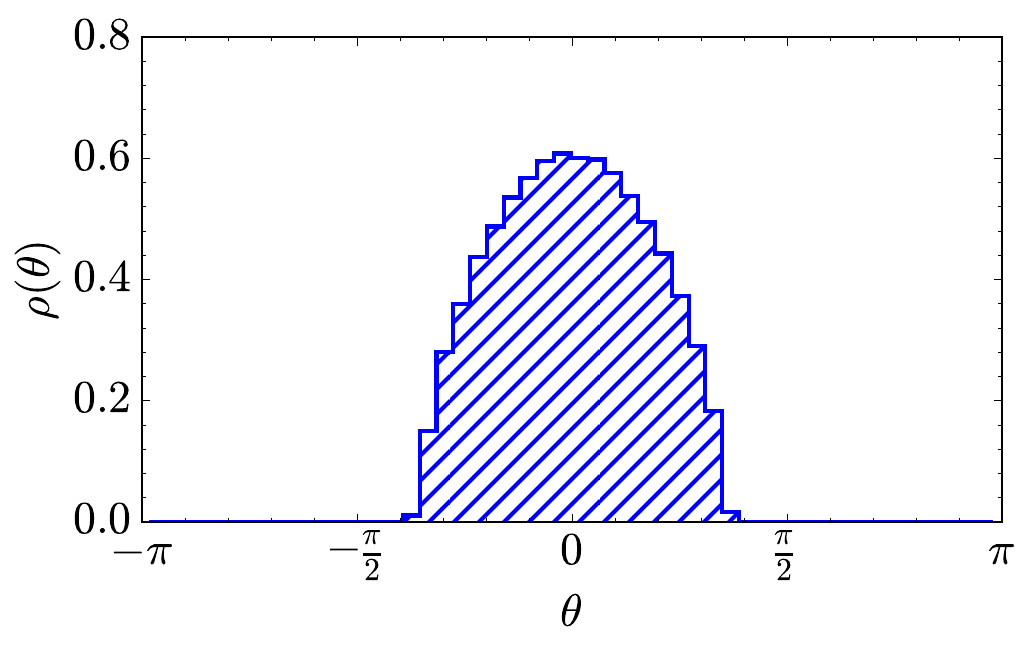}
   \caption{The Polyakov loop scatter plot (top) and its eigenvalue distribution (bottom) with $\hat \mu = 2, N = 32$ for temperatures $\hat T = 0.8$ (red), $0.913$ (green) and $1.3$ (blue). The eigenvalue distribution for these temperatures corresponds to uniform, non-uniform, and gapped phases, respectively.}
  \label{fig:dist}
\end{figure}

\section{Phase Diagram}

\begin{figure}[h!]
  \centering
    \includegraphics[width=0.7\linewidth]{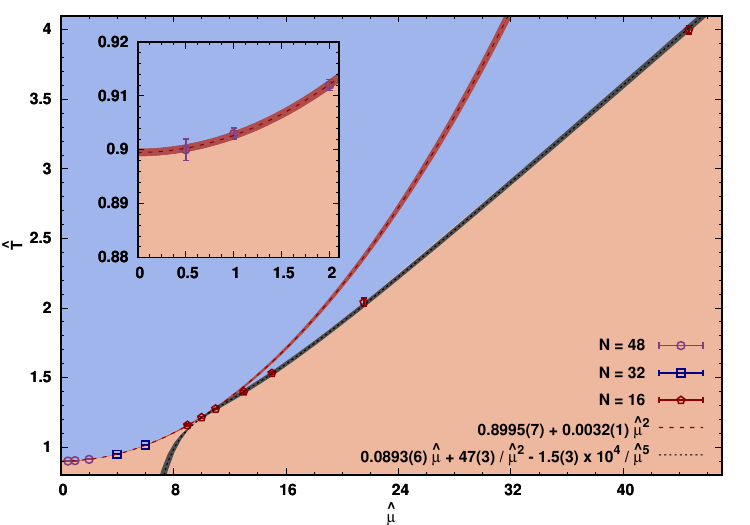}
  \caption{The $\hat T$ - $\hat \mu$ phase diagram of the bosonic BMN model from our $N_\tau = 24$ results. The blue-shaded region in the diagram represents the deconfined phase, and the red-shaded region represents the confined phase.}
  \label{fig:phase}
\end{figure}

In Fig. \ref{fig:dist}(top), we show the Polyakov loop scatter plots for a fixed deformation mass $\mu = 2.0$ for different temperatures. The corresponding eigenvalue distributions of the Polyakov loop are also shown in Fig. \ref{fig:dist}(bottom). For different values of the deformation parameter used, we calculated the transition point. In Fig.~\ref{fig:phase}, we show the corresponding phase diagram.
This phase diagram smoothly interpolates between the limits of the bosonic BFSS model and the gauged Gaussian model with first order phase transition for all couplings, $g = \lambda/\mu^3$. The comprehensive analysis of this work can be found in Ref.~\cite{Dhindsa:2022uqn}. The results for the critical temperature over extreme coupling regimes ($g = 0$ and $g \rightarrow \infty$) match with previous studies \cite{Aharony:2003sx,Bergner:2021goh}, and our analysis shows that these extreme regimes can be connected using a smooth function.

\section{Future Directions}
We plan to continue exploring the strong coupling regime ($g \rightarrow \infty$) with the larger values of $N$ to verify the results with other studies. We also plan to study the ungauged versions of the bosonic and the full BMN models.  

\section{Acknowledgments}
We thank Raghav Jha for fruitful discussions and collaboration on this work. NSD thanks the Council of Scientific and Industrial Research (CSIR), Government of India, for the financial support through a research fellowship (Award No. 09/947(0119)/2019-EMR-I). The work of AS was
partially supported by an INSPIRE Scholarship for Higher Education by the Department of Science and Technology, Government of India. AJ was supported in part by IISER Mohali and the University of the Witwatersrand. DS was supported by UK Research and Innovation Future Leader Fellowship MR/S015418/1 and STFC grant ST/T000988/1. Numerical calculations were carried out at the University of Liverpool and IISER Mohali.

\printbibliography

\end{document}